\documentclass[12pt]{article}
\usepackage{graphicx}
\usepackage{amsmath}

\begin{document}

\title{A middle option for choices in the Continuous Opinions and Discrete Actions model}
\author{\underline{Andr\'e C. R. Martins}\\%, 2. Author$^{3}$ \\
%EndAName
GRIFE - EACH, Universidade de S\~ao Paulo, Brazil\\
(amartins@usp.br) }
\date{}
\maketitle

\begin{abstract}
Modeling the conditions for the emergence of extremism is a very important problem, with clear applications for describing the interaction among individuals. Traditional models either are not suited for the task, as in the case of discrete models, or, like Bounded Confidence models, are built with rules that make opinions tend to a common ground between agents or not change at all. Continuous Opinions and Discrete Actions (CODA) model allowed us to observe the emergence of extremist agents, even when every agent was initially a moderate, due to local influence effects. In this paper, the problem of emergence of extremism will be addressed by introducing a middle discrete option in the CODA model, making it similar to a Potts model. Different scenarios for the third option will be discussed: when it is equivalent to withholding judgment, when it is a real third option and when it is a real, middle option. The effects on the opinions will be studied and its effects on extremism discussed. Withholding judgment seems to have an unexpected effect, causing the diminishing of moderate opinions in the long run. For a central third opinion, we find that, under specific conditions, this new choice can act as a buffer between the extreme choices.
%\medskip \noindent {Key Words: } {\ Opinion Dynamics, Sociophysics, Extremism, CODA, Bayesian update rules}
\end{abstract}

\section{Introduction}

In order to describe the social evolution of the strength of opinions about an issue, models that allow for a large number of possible internal opinions are needed, or even continuous opinions, as in the Bounded Confidence models \cite{deffuantetal00,hegselmannkrause02}. However, those models only use tendency towards consensus. Although they can be useful at describing the spread of an already existing extremist position \cite{deffuantetal02a,amblarddeffuant04,deffuant06,gargiulomazzoni08a}, they are not capable of explaining the emergence of extreme opinions. 

On the other hand, people often express themselves by choosing from one of a small number of options. This choice aspect is well captured by the discrete models \cite{galametal82,galammoscovici91,sznajd00,stauffer03a}, but at the cost of measuring the strength of the opinions. While one can introduce in any of these models the difference between opinion and verbalization \cite{urbig03}, it is more natural to introduce models based on Bayesian Decision Theory \cite{martins08e}, where the verbalized choices and inner probabilistic opinions are naturally defined. The application of that idea to discrete choices lead to the development of the Continuous Opinions and Discrete Actions (CODA) model \cite{martins08a,martins08b} where the emergence of extremism was observed as a consequence of the local reinforcement of the opinions. An extension of those ideas using perceptrons in a cultural environment where many issues are debated also showed regimes where extremism arises naturally \cite{vicenteetal08b}. Bayesian update rules was also used for continuous verbalized information and results similar to those observed in the Bounded Confidence models were obtained as a particular case \cite{martins08c}. While no emergence of extremism was observed, when agents updated also their uncertainty, stubbornness was observed as a consequence of the rules.

The discrete choices used in the CODA model, however, were only binary choices, represented as a spin associated to each agent $i$, $s_i=\pm1$. Therefore, it makes sense to explore the consequences for the dynamics of the system and, in particular, the emergence of extremism when we introduce a third choice in the problem \cite{gekleetal05a,voloviketal09a,hsuhuang09a}. 
Two cases will be explored in this paper. First, we will see that the binary CODA model allows for the easy introduction of a middle, agnostic, third choice. This corresponds to the case where no option is made between two real choices when the agent is not really sure and, therefore, it makes no claims about its preferred option. The second case we will study corresponds to the introduction of a real third choice. We will investigate both the case where the third option is independent of the other two and where it is a middle choice between two extremes.

\section{Witholding Judgement in the CODA model}

Assume there are two real choices $s_i=\pm 1$ agent $i$ can make, associated with a decision. In order to evaluate which choice is better, $i$ has an internal subjective probability, $p_i$ associated with the possibility that the choice $s_i =+1$ is the best one by $s_i=sign(p_i -0.5)$. If the agent believes its neighbors are mostly like right than wrong, that is, each neighbor will choose the best option with probability $\alpha>0.5$, then, when observing the choices of a neighbor $j$, agent $i$ should update its internal probability using Bayes Theorem. 

It was shown previously that by using the transformed variable $\nu_i=\ln(p_i/1-p+i)$, the problem becomes much simpler. Each time $s_j=+1$ is observed the log-odds opinion $\nu_i$ increases by a fixed amount and, when $s_j=-1$ is observed, $\nu_i$ decreases by the same amount. Therefore, by measuring $\nu_i$ in number of steps, we don't even have to worry about the step size (this will no longer be true for three-choice problems).

\begin{figure}[ht]
 \includegraphics[width=0.8\textwidth]{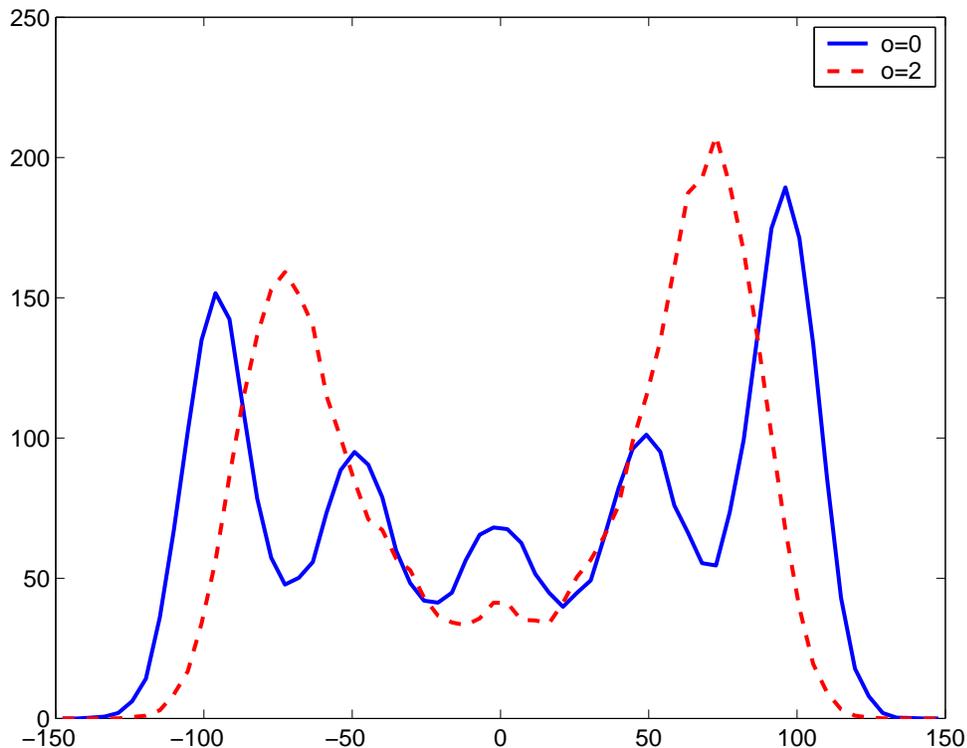}
 \caption{Distribution of the opinions, measured in number of steps away from changing the choice, when the middle choice is not a real choice, but just a statement of ignorance. The cases where $o=0$ (no ignorance) and $o=2$ are both shown. In both cases the intitial conditions where chosen so that the initial opinions were distributed uniformly in a range of 2.025 steps away from $\nu=0$.
}\label{fig:opdistn64d2}
\end{figure}

It is easy to introduce agents who withhold judgment in this context. Since $p_i=0.5$ translates to $\nu_i=0$, instead of simply assigning $s_i=sign(\nu_i)$, we can have a region of size $o$ around $\nu_i=0$, where each agent express no opinion and, therefore, has no influence on its neighbors. That is, if $|\nu_i|<o$, then agent $i$ withholds judgment until it is less uncertain about the problem.

Figure \ref{fig:opdistn64d2} shows the opinion distributions, measured in number of steps away from changing the choice, after an average of 100 interactions per agent. The agents were located in a bi-dimensional $64x64$ square regular lattice with four neighbors for each agent. The figure shows the cases where $o=0$ (no ignorance) and $o=2$ and the initial conditions where chosen so that the initial opinions were distributed uniformly in a range of 2.025 steps away from $\nu_i=0$. This means that, in the $o=2$ case, few agents will have a stated opinion in the beginning. Since a neighbor with no stated opinion has no influence, the system evolves slowly at first, while small clusters of likely-minded individuals spread until most agents have a very strong opinion on the subject. This suggests that, while extremism has a slower start, once most people express one real option, the behavior of the system should be similar to that of the model with no agnostic agents.

\begin{figure}[htp]
\hspace{-0.5cm}\includegraphics[width=0.4\textwidth]{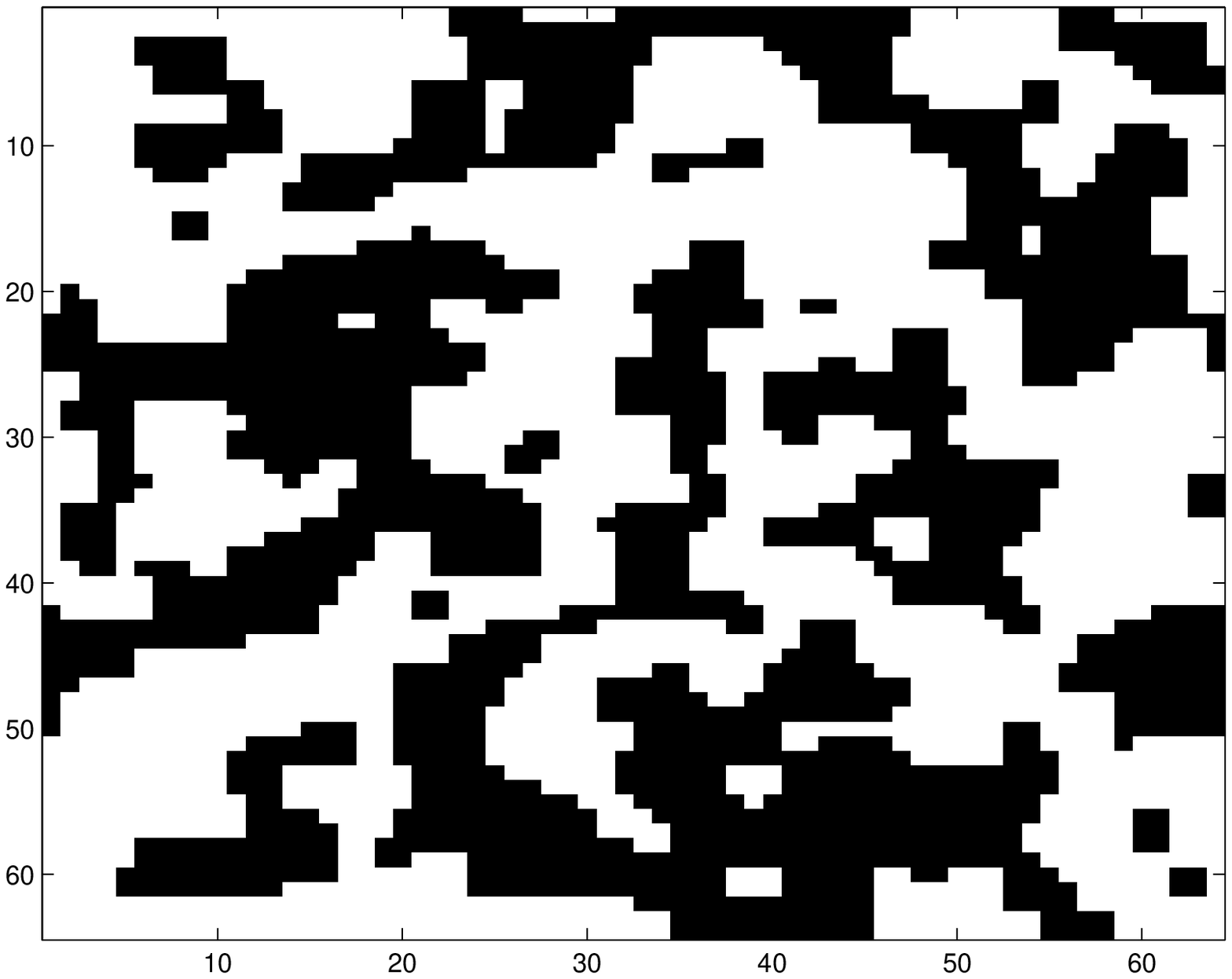}\hspace{0.5cm}
\includegraphics[width=0.4\textwidth]{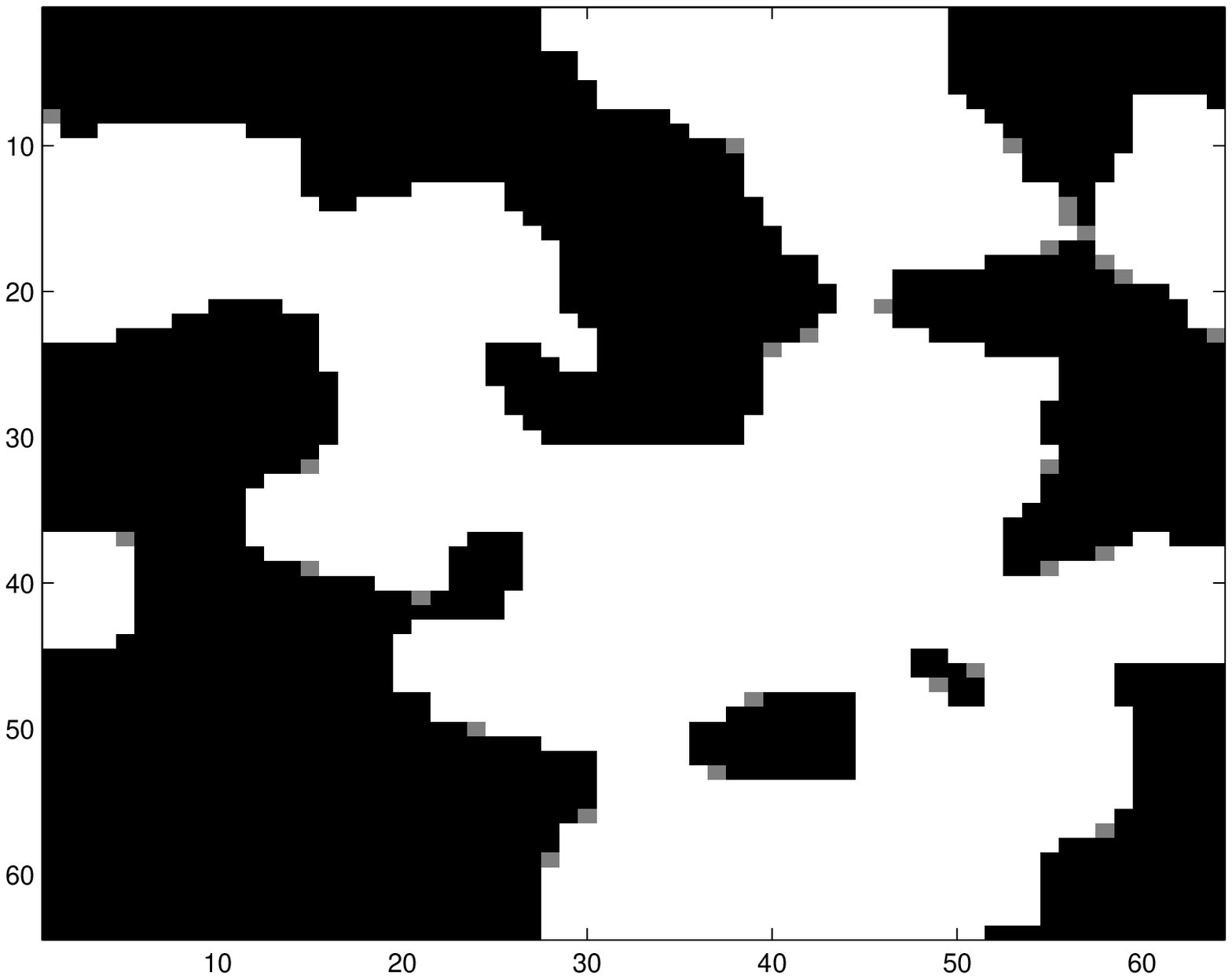}
 \caption{Configuration for the regular square lattice with 64x64 agents when agents can withhold judgment until they are certain (grey squares show agents who are not certain). (\textit{Left panel}) No ignorance ($o=0$) (\textit{Right panel}) Most agents start with no chosen position ($o=2$)}\label{fig:configop}
 \end{figure}
 
However, while there is only a small change in the strength of the extremism, the final configuration of the system shows important differences. Figure \ref{fig:configop} shows the final configuration for both cases presented in Figure \ref{fig:opdistn64d2}. We see that for large $o$, the regions of individuals that make the same choice become larger and more agents find themselves in the middle of areas where no dissent is observed. This leads to an important decrease in the number of agents that live in interfaces between the regions. As a consequence, the central peaks become less important. That is, while the final extremism is a little weaker, we observe less agents with moderate opinions. The initial doubt has the counter-intuitive effect of decreasing doubt in the long run, as clusters become larger and the influence between the groups, weaker.

\section{Three real choices}

To introduce a third real choice, we need to alter the CODA model. We will introduce three choices, $A$, $B$ and $C$. When an agent $i$ chooses $A$, it sets its choice at $s_i=-1$; for $B$, $s_i=0$; and, for $C$, $s_i=+1$. In particular, with three choices, the agents need, in the most general case, estimates of six different conditional probabilities (the nine chances that each result will be picked given each result is the best less the three normalization equations). In order to make the problem simpler, we can introduce a few symmetries. Therefore, we will assume that the chances agents assign to the possibility that each neighbor will choose the best option is the same for all three options. That is, 
\[
P(s_j=-1|A)=P(s_j=-0|B)=P(s_j=+1|C)=a.
\] 
Also, by introducing a symmetry between choices $A$ and $C$, we will have that  $P(s_j=+1|B)= P(s_j=-1|B)=(1-a)/2$ (since it must adds to one) and 
\[
P(s_j=0|A) = P(s_j=0|C)=b.
\]
This reduces the number of independent likelihoods to two.

The initial probabilities associated with $A$, $B$, and $C$, assigned by agent $i$ will be represented by $p_i$, $q_i$, and $r_i$, respectively. In order to simplify the notation, the index $i$ will not be used from now on. Changing the probability variables for $\ln(p/(1-p)$ no longer simplifies the dynamical equations one obtains from a direct application of Bayes Theorem. However, if we define
\[
\nu_1=\ln\left( \frac{p}{q}\right) ,
\]
\[
\nu_2=\ln\left( \frac{q}{r}\right) ,
\]
\begin{equation}\label{eq:definenu}
\nu_3=\ln\left( \frac{r}{p} \right) ,
\end{equation}
where, obviously, $\nu_1+\nu_2+\nu_3=0$, we have a set of variables where the process of opinion change by the use of the Bayes Theorem is still additive. We have that, when the choice $s_j=-1$ is observed in the neighbor, the variables transform according to
\[
\nu_1(t+1)=\nu_1(t)+\ln\left( \frac{2a}{1-a}\right),
\]
\begin{equation}
\nu_2(t+1)=\nu_2(t)+\ln\left( \frac{1-a}{2(1-a-b)}\right),
\end{equation}
where the equation for $\nu_3$ was omitted since it is a direct consequence of the previous pair. If $s_j=0$ is observed, we have
\[
\nu_1(t+1)=\nu_1(t)+\ln\left( \frac{b}{a}\right),
\]
\begin{equation}
\nu_2(t+1)=\nu_2(t)+\ln\left( \frac{a}{b}\right),
\end{equation}
and, finally, for $s_j=+1$,
\[
\nu_1(t+1)=\nu_1(t)+\ln\left( \frac{2(1-a-b)}{1-a}\right),
\]
\begin{equation}\label{eq:stepsizes}
\nu_2(t+1)=\nu_2(t)+\ln\left( \frac{1-a}{2a}\right).
\end{equation}

From Equation \ref{eq:definenu} we see that the sign of the variables $\nu_1$, $\nu_2$ and $\nu_3$ define which probability in the pair used to calculate the variables is larger. Therefore, simply by checking the signs, it is possible to determine which probability, $p$, $q$, or $r$, is considered larger by the agent and, therefore, the choice the agent will express. 

The problem with the use of variables $\nu_1$, $\nu_2$ and $\nu_3$ is that, while they simplify the dynamics considerably, they are not related to just one specific choice. Therefore, in order to interpret the data we will obtain, they need to be translated back to probabilities or true log-odds in favor of each choice. By solving the Equation \ref{eq:definenu}, we get
\[
l_1=\ln \left( \frac{p}{1-p} \right) = \nu_1-\ln \left( e^{-\nu_2}+1 \right),
\]
\[
l_2=\ln \left( \frac{q}{1-q} \right) = \ln \left( \frac{1}{e^{\nu_1}+e^{-\nu_2}} \right),
\]
\begin{equation}\label{eq:logodds}
l_3=\ln \left( \frac{r}{1-r} \right) = -\nu_2+\ln \left( e^{\nu_1}+1 \right).
\end{equation}
If necessary, one can compute the probabilities directly from the results of Equation \ref{eq:logodds}, but the values of $l_1$, $l_2$, and $l_3$ are enough to understand the consequences for the system. Actually, since extreme opinions might evolve, turning the probabilities too close to 0 and 1, they are actually a better choice of variables than $p$, $q$, and $r$.

\subsection{Simulations}

 In order to explore the consequences of introducing a third option in the problem, a series of simulations was performed. All cases were run in a square bi-dimensional regular lattice with periodic boundary conditions and interactions with only the four closest neighbors (von Neumann neighborhood). The initial conditions were chosen by drawing random numbers uniformly distributed in the interval $1/3\pm 0.1$ for each choice. Once the sum was renormalized to 1, those numbers were used as initial values for $p$, $q$, and $r$. The largest probability was the criterion for picking each agent choice. That is, at first, there were no agents with extreme opinions.

\begin{figure}[htp]
\hspace{-0.5cm}\includegraphics[width=0.4\textwidth]{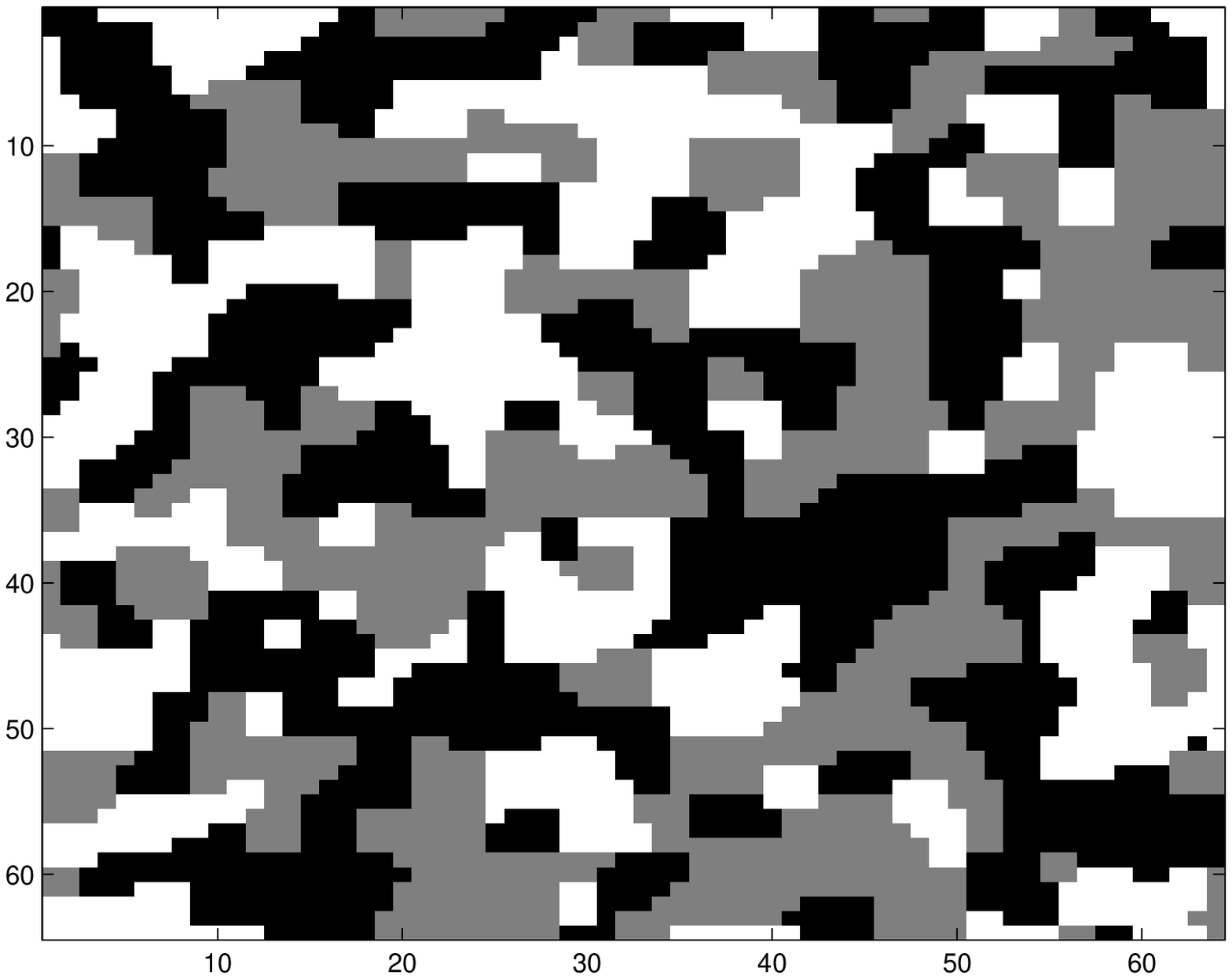}\hspace{0.5cm}
\includegraphics[width=0.4\textwidth]{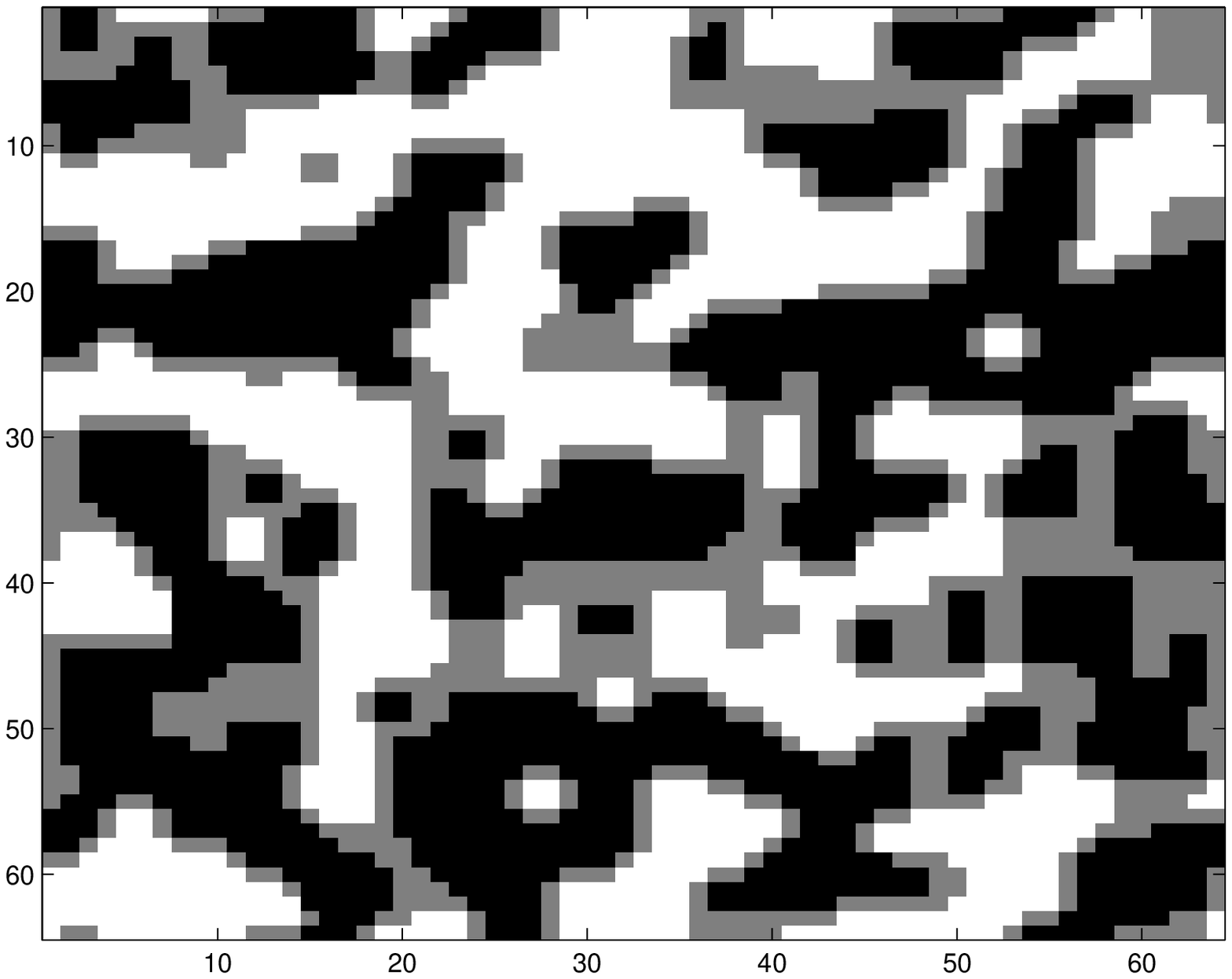}
 \caption{Configuration for the regular square lattice with 64x64 agents when there are three real choices. In both cases $\alpha=0.7$. (\textit{Left panel}) The $s_i=0$ choice (grey) is identical as the others and not a middle option ($\beta=(1-\alpha)/2$=0.15). (\textit{Right panel}) The $s_i=0$ choice is a middle choice and $s_i=1$ and $s_i=-1$ are considered very distant ($\beta=0.29$).}\label{fig:confpott}
 \end{figure}
 
Figure \ref{fig:confpott} shows the stable lattice configuration of choices once after one typical realization for each case shown. In the figure, white means the agent at that site chose $s_i=-1$, grey means the middle choice, $s_i=0$, and black corresponds to $s_i=+1$. Configurations for the case where the middle choice is an independent third option ($\beta=(1-\alpha)/2$=0.15), as well as when it is a real middle choice ($\beta=0.29$) are shown. It is easy to see that there are some important differences between both figures. Notice that $\beta=0.29$ is a case where each agent thinks it is very unlikely (only 1\%) that a neighbor will make an extreme choice when the opposite choice is the best one (the neighbor $j$ choosing $s_j=+1$, when $A$ is the best choice, per example). When that happens, we can see that the middle choice actually survives mostly as a buffer between the white and black areas, as opposed to the case where the grey choice was independent and behaved in the same way as the other two.

\begin{figure}[htp]
\hspace{-0.5cm}\includegraphics[width=0.45\textwidth]{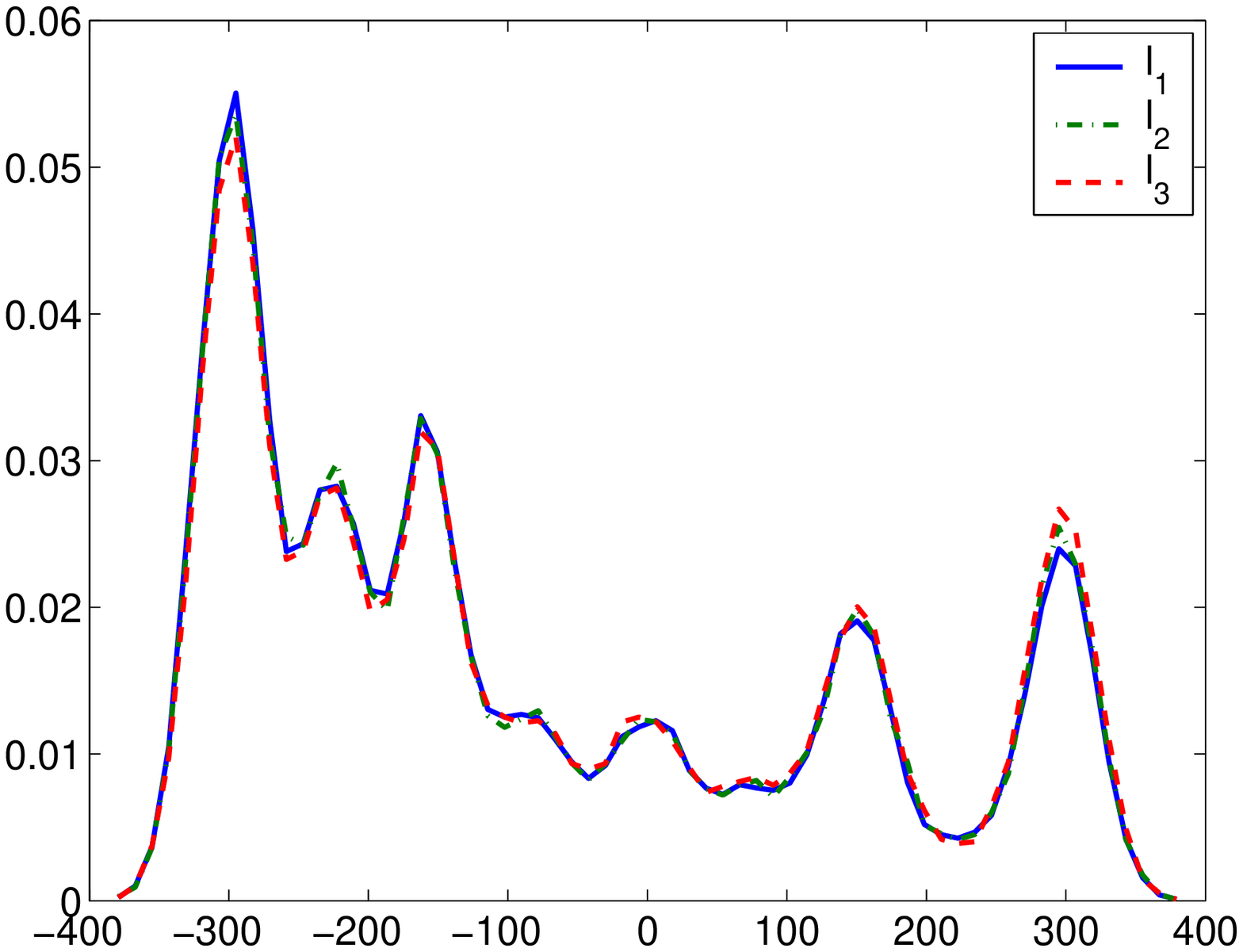}\hspace{0.5cm}
\includegraphics[width=0.45\textwidth]{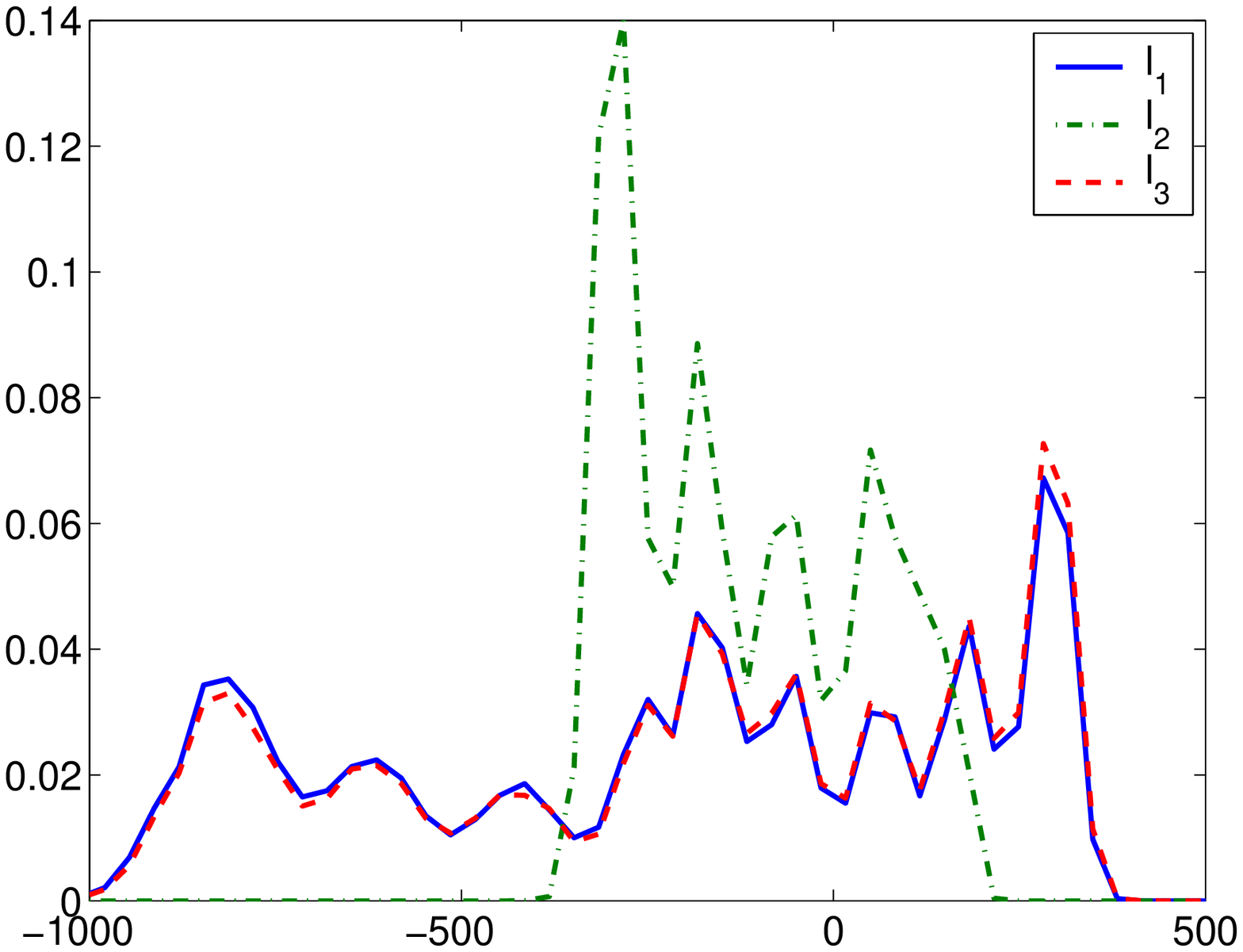}
 \caption{Opinion distribution for each choice in the regular square lattice with 64x64 agents when there are three real choices. In both cases $\alpha=0.7$. (\textit{Left panel}) The $s_i=0$ choice is identical as the others and not a middle option ($\beta=(1-\alpha)/2$=0.15). (\textit{Right panel}) The $s_0$ choice is a middle choice and $s_i=+1$ and $s_i=-1$ are considered very distant ($\beta=0.29$).}\label{fig:distpotts}
 \end{figure}

The strength of the opinions is shown in Figure \ref{fig:distpotts}, for the same case as in Figure \ref{fig:confpott}. Here, the histograms were obtained as average distributions after 20 realizations of the system. They show the distribution for the log-odds of each log-odds variable associated with each choice, that is,  $l_1$, $l_2$, and $l_3$. As expected, when the $s_0$ choice is an independent third choice, all three histograms present the same behavior. We can see the appearance of peaks in the most extreme positions, both for positive $l$ values (agents who choose that specific option), as well as negative (agents who think that the evaluated option is very unlikely to be the best one). As the problem is symmetrical regarding the three options, it is expected that, in average, each option should attract one third of the choices. This is what makes the peak of the agents who believe a theory is wrong approximately twice as high as the peak of agents who think it is right. 

The $\beta=0.29$ case shows the disagreement extreme peak much further than the agreement peak for both $l_1$ and $l_3$. This has happened because whenever an agent with a non-intermediary choice gets its opinion reinforced, the step resulting from Equation \ref{eq:stepsizes} against the opposite choice is larger than the reinforcements. Therefore, the opinion against the opposite opinion ($s_i=-1$ or $s_i=+1$) ends larger than the opinion in favor of any of the choices. It is interesting to notice that the opinion against the central option is never so strong as those against the more non-central choices.

\begin{figure}[htp]
\hspace{-0.5cm}\includegraphics[width=0.45\textwidth]{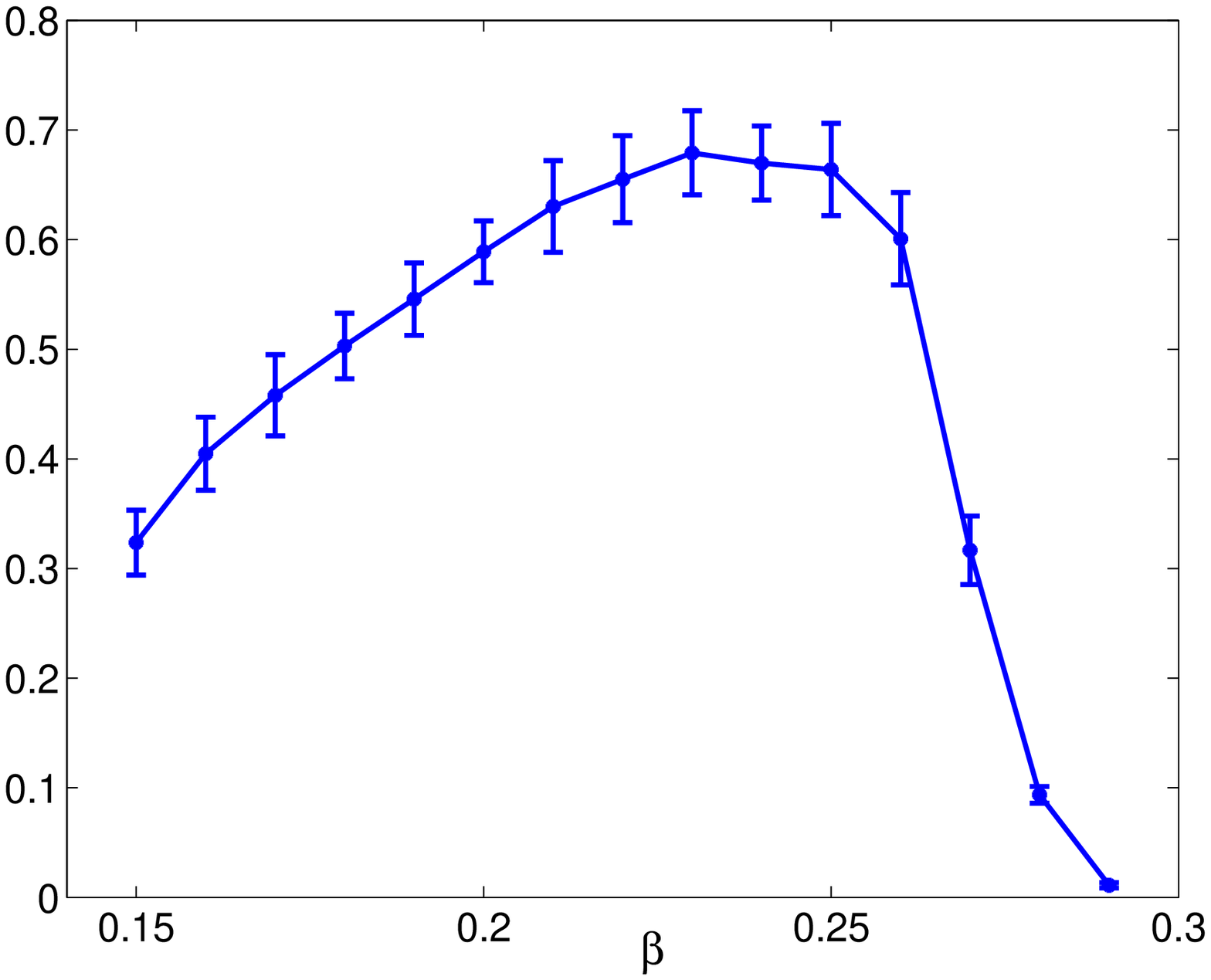}\hspace{0.5cm}
\includegraphics[width=0.45\textwidth]{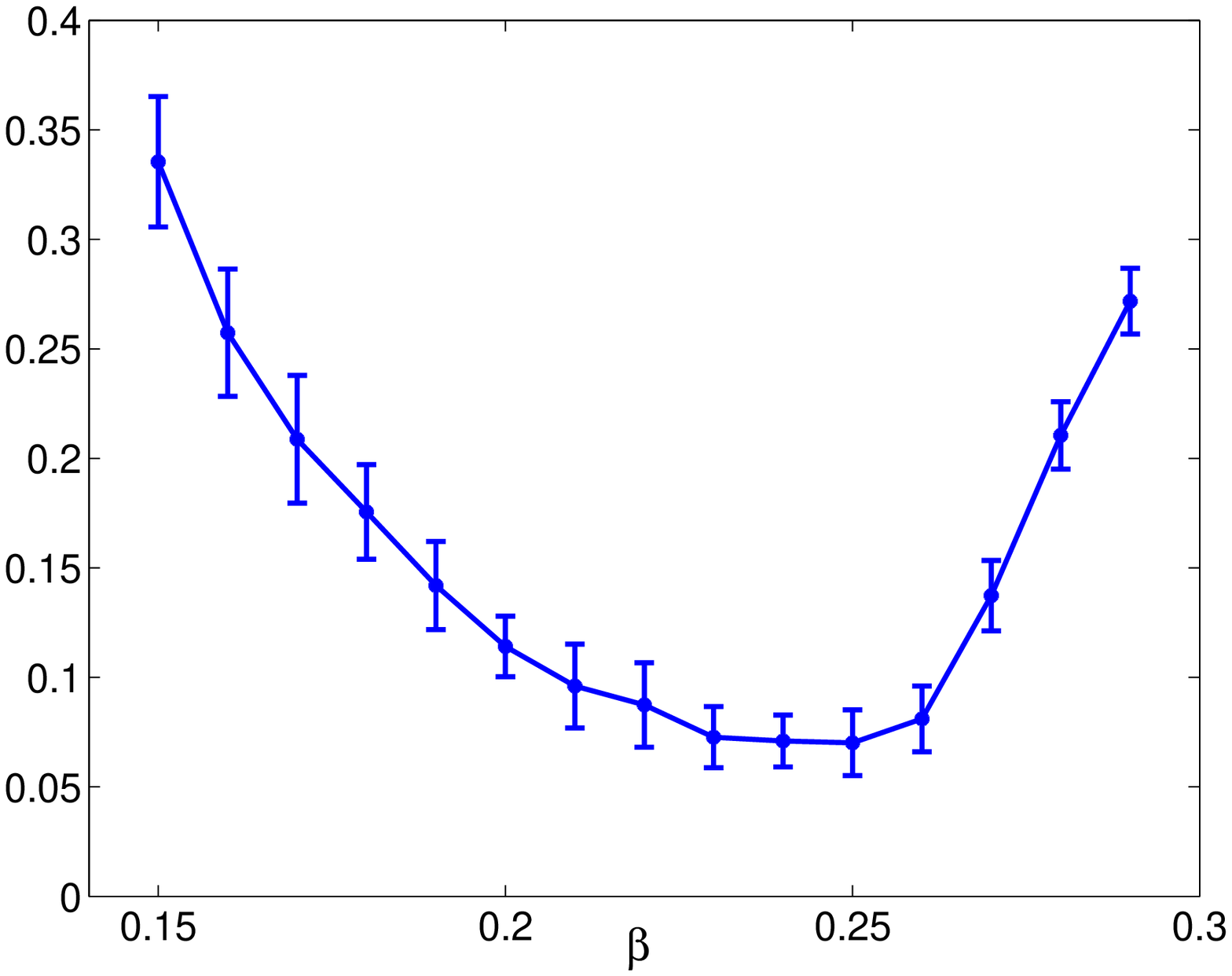}
 \caption{Effects on the configuration of changing the value of $\beta$, while keeping $\alpha=0.7$ constant. Each point shows the average result of 20 realizations and the error bars are the observed standard deviations. (\textit{Left panel}) The proportion of interfaces between $s_i=+1$ and $s_i=-1$, relative to the total number of interfaces. (\textit{Right panel}) Proportion of $s_i=0$ results that survive at the end.}\label{fig:interfaces}
 \end{figure}

Finally, Figure \ref{fig:interfaces} investigates the interfaces between the different opinions, as we change $\beta$ from $\beta=0.15$ to $\beta=0.29$, including the intermediary values, while keeping $\alpha=0.7$. The estimates shown are the averages of 20 different realizations and the error bars were obtained from the standard deviation of the quantities measured. The left panel shows the proportion of interfaces between $s_i=+1$ and $s_i=-1$, relative to the total number of interfaces. Since $s_i=+1$ and $s_i=-1$ are symmetric, their average number of interfaces with $s_0$ are equal and, therefore, do not need to be evaluated. We see that, contrary to what we would expect from the analysis from just the two extreme cases ($\beta=0.15$ and $\beta=0.29$), the proportion of interfaces between the opposing factions increase at first and only starts decreasing after $\beta=0.25$.

The reason for this behavior can be understood in the right panel. We see that, as $\beta$ increases, the proportion of agents who choose the middle option decreases at first; around $\beta=0.25$, only as little as 8\% of agents keep the intermediary position. Since there are so few of them, it is natural that most interfaces will not include agents with the $s_i=0$ choice. However, as $\beta$ increases even further, this tendency is reversed and the number of agents who choose the middle option becomes larger. They survive in the areas between the opposing choices $s_i=+1$ and $s_i=-1$. Simulations for $\beta=0.299$ show that the interfaces between the opposite choices actually disappear, in 20 realizations not even one case of neighbors between $s_i=+1$ and $s_i=-1$ was observed. 

\section{Conclusions}

We see that, as $s_i=0$ becomes a central option, initially, this option loses its appeal and is replaced by the two competing extremes. However, if the extremes are such that, if one of them is true, it is considered very unlikely that a neighbor would choose the opposite option, the central opinion regain its strength and becomes a buffer between the $s_i=+1$ and $s_i=-1$. Opinions about each option still remain strong with the introduction of the third option, as shown by the histograms of the log-odds, but, from a practical point of view, the existence of agents that have the intermediary opinion acting as buffers could cause a decrease in extremist problems. It is interesting to see that the middle option always survive, although it becomes much less important for a range of values.

The introduction of a third option in a binary choice problem, by allowing agents who are not very sure to not express their opinions until those opinions are strong enough also had interesting consequences. While a system where most agents have no chosen factions at first leads to less extreme opinions, due to the fact that for a while there is no reinforcement while most agents express no choice, the overall effect can not be described as a diminishing of extremism. As a matter of fact, with few agents with an initial opinion, the domains can grow larger before the system dynamics freeze them. This has the effect that a smaller number of agents live in the interfaces and observe both choices. Therefore, the number of moderates actually decreases as an effect of the agents withholding their opinions when they are not so certain.

\section*{Acknowledgement}

The author would like to thank Funda\c{c}\~ao de Amparo \`a  Pesquisa do Estado de S\~aoPaulo (FAPESP) for the support to this work, under grant 2008/00383-9.

 \bibliographystyle{unsrt}
 \bibliography{biblio}

\begin{thebibliography}{10}

\bibitem{deffuantetal00}
G.~Deffuant, D.~Neau, F.~Amblard, and G.~Weisbuch.
\newblock Mixing beliefs among interacting agents.
\newblock {\em Adv. Compl. Sys.}, 3:87--98, 2000.

\bibitem{hegselmannkrause02}
R.~Hegselmann and U.~Krause.
\newblock Opinion dynamics and bounded confidence models, analysis and
  simulation.
\newblock {\em Journal of Artificial Societies and Social Simulations}, 5(3):3,
  2002.

\bibitem{deffuantetal02a}
G.~Deffuant, F.~Amblard, and T.~Weisbuch, G.and~Faure.
\newblock How can extremism prevail? a study based on the relative agreement
  interaction model.
\newblock {\em JASSS-The Journal Of Artificial Societies And Social
  Simulation}, 5(4):1, 2002.

\bibitem{amblarddeffuant04}
F.~Amblard and G.~Deffuant.
\newblock The role of network topology on extremism propagation with the
  relative agreement opinion dynamics.
\newblock {\em Physica A}, 343:725--738, 2004.

\bibitem{deffuant06}
G.~Deffuant.
\newblock Comparing extremism propagation patterns in continuous opinion
  models.
\newblock {\em JASSS-The Journal Of Artificial Societies And Social
  Simulation}, 9(3):8, 2006.

\bibitem{gargiulomazzoni08a}
Floriana Gargiulo and Alberto Mazzoni.
\newblock Can extremism guarantee pluralism?
\newblock {\em JASSS-The Journal Of Artificial Societies And Social
  Simulation}, 11(4):9, 2008.

\bibitem{galametal82}
S.~Galam, Y.~Gefen, and Y.~Shapir.
\newblock Sociophysics: A new approach of sociological collective behavior:
  Mean-behavior description of a strike.
\newblock {\em J. Math. Sociol.}, 9:1--13, 1982.

\bibitem{galammoscovici91}
S.~Galam and S.~Moscovici.
\newblock Towards a theory of collective phenomena: Consensus and attitude
  changes in groups.
\newblock {\em Eur. J. Soc. Psychol.}, 21:49--74, 1991.

\bibitem{sznajd00}
K.~Sznajd-Weron and J.~Sznajd.
\newblock Opinion evolution in a closed community.
\newblock {\em Int. J. Mod. Phys. C}, 11:1157, 2000.

\bibitem{stauffer03a}
D.~Stauffer.
\newblock How to convince others? monte carlo simulations of the sznajd model.
\newblock In J.~E. Gubernatis, editor, {\em AIP Conf. Proc. v. 690: The Monte
  Carlo Method in the Physical Sciences: Celebrating the 50th Anniversary of
  the Metropolis Algorithm}, pages 147--155. American Institute of Physics,
  2003.

\bibitem{urbig03}
D.~Urbig.
\newblock Attitude dynamics with limited verbalisation capabilities.
\newblock {\em Journal of Artificial Societies and Social Simulations}, 6(1):2,
  2003.

\bibitem{martins08e}
Andr\'e C.~R. Martins.
\newblock A bayesian framework for opinion updates.
\newblock arXiv:0811.0113v1, 2008.

\bibitem{martins08a}
Andr\'e C.~R. Martins.
\newblock Continuous opinions and discrete actions in opinion dynamics
  problems.
\newblock {\em Int. J. of Mod. Phys. C}, 19(4):617--624, 2008.

\bibitem{martins08b}
Andr\'e C.~R. Martins.
\newblock Mobility and social network effects on extremist opinions.
\newblock {\em Phys. Rev. E}, 78:036104, 2008.

\bibitem{vicenteetal08b}
R.~Vicente, Andr\'e C.~R. Martins, and N.~Caticha.
\newblock Opinion dynamics of learning agents: Does seeking consensus lead to
  disagreement?
\newblock {\em Journal of Statistical Mechanics: Theory and Experiment},
  2009:P03015, 2009.
\newblock arXiv:0811.2099.

\bibitem{martins08c}
Andr\'e C.~R. Martins.
\newblock Bayesian updating rules in continuous opinion dynamics models.
\newblock {\em Journal of Statistical Mechanics: Theory and Experiment},
  2009(02):P02017, 2009.
\newblock arXiv:0807.4972v1.

\bibitem{gekleetal05a}
S.~Gekle, L.~Peliti, and S.~Galam.
\newblock Opinion dynamics in a three-choice system.
\newblock {\em Eur. Phys. J. B}, 45:569--585, 2005.

\bibitem{voloviketal09a}
D.~Volovik, M.~Mobilia, and S.~Redner.
\newblock Dynamics of strategic three-choice voting.
\newblock {\em EPL}, 85:48003, 2009.

\bibitem{hsuhuang09a}
Jiann-Wien Hsu and Ding-Wei Huang.
\newblock Local effects lead to uncertainty in opinion dynamics.
\newblock {\em Int. J. Mod. Phys. C}, 20(1):9--23, 2009.

\end{thebibliography}

\end{document}